\documentclass[twocolumn,floats,showpacs,superscriptaddress,pre]{revtex4}
\usepackage{amsmath,amssymb,bm}
\usepackage{graphics}
\usepackage{epsfig}
\usepackage{graphicx}

\begin{document}

\title{On a Quantum Action Principle}
\author{Natalya Gorobey and Alexander Lukyanenko}
\email{alex.lukyan@rambler.ru}
\affiliation{Department of Experimental Physics, St. Petersburg State Polytechnical
University, Polytekhnicheskaya 29, 195251, St. Petersburg, Russia}

\begin{abstract}
A quantum version of the action principle is formulated in terms of real
parameters of a wave functional. The classical limit of the quantum action
of a harmonic oscillator is obtained.
\end{abstract}

\maketitle
\date{\today }





\section{INTRODUCTION}

In the work \cite{GL} a new form of quantum mechanics in terms of a
quantum version of the action principle was proposed. The quantum
action principle was formulated for a new object - a wave functional
$\Psi \left[ x\left( t\right) \right] $, which, unlike a wave
function $\psi \left( x,t\right) $, describes the dynamics of a
particle as the movement along a trajectory $x\left( t\right) $. The
wave functional has the meaning of a probability density in the
space of trajectories, provided that a scalar product in the space
of functionals is introduced and the normalization condition

\begin{equation}
\left( \Psi ,\Psi \right) \equiv \int \prod_{t}dx(t) \left\vert \Psi
\left[ x\left( t\right) \right] \right\vert ^{2}=1 \label{1}
\end{equation}%
is fulfilled. The new form of quantum mechanics is equivalent to
Schr\"{o}dinger wave mechanics on a set of multiplicative wave
functionals that are connected with an $\varepsilon $-division of a
time interval $\left[ 0,T \right] $ \cite{GL}.

In the present paper the quantum action principle is formulated as
an extremum principle on a set of analytical functionals, i.e. the
functionals that are represented by a functional series of $x\left(
t\right) $ degrees. In this framework the dynamics of a particle is
reduced to a system of differential equations for coefficients of
the series. We consider quantum dynamics of a harmonic oscillator
and obtain its quasi-classical limit.

\section{QUANTUM ACTION PRINCIPLE}

The quantum action principle is based on the following equation for
eigenvalues of the action operator:

\begin{eqnarray}
\widehat{I}\Psi &\equiv &\int_{0}^{T}dt\left[ \frac{\widetilde{\hbar }}{i}%
\overset{\cdot }{x}\left( t\right) \frac{\delta \Psi }{\delta x\left(
t\right) }+\frac{\widetilde{\hbar }^{2}}{2m}\frac{\delta ^{2}\Psi }{\delta
x^{2}\left( t\right) }\right.  \notag \\
&&\left. -U\left( x\left( t\right) ,t\right) \Psi \right] \left. =\right.
\left. \lambda \Psi \right. .  \label{2}
\end{eqnarray}%
The action operator $\widehat{I}$ is a quantum analog of the classical
action represented in the canonical form:

\begin{equation}
I=\int_{0}^{T}dt\left[ p\overset{\cdot }{x}-\frac{p^{2}}{2m}-U\left(
x,t\right) \right] .  \label{3}
\end{equation}%
"Quantization" of the classical action (\ref{3}) is performed by the
replacement of the canonical momentum $p\left( t\right) $ by the
functional-differential operator

\begin{equation}
\widehat{p}\left( t\right) \equiv \frac{\widetilde{\hbar }}{i}\frac{\delta }{%
\delta x\left( t\right) },  \label{4}
\end{equation}%
where the constant $\widetilde{\hbar }$ has the dimensionality $Dj\cdot
s^{2} $. On the set of multiplicative functionals, which corresponds to an $%
\varepsilon $-division of the time interval $\left[ 0,T\right] $, this
constant is proportional to the "ordinary" Plank constant:

\begin{equation}
\widetilde{\hbar }=\hbar \cdot \varepsilon .  \label{5}
\end{equation}

For the set of analytical functionals considered here the constant $%
\widetilde{\hbar }$ has to be defined after a definition of
"observables" and a comparison of the theory with an experiment.
This will be the subject of our next work.

Let us suppose that a potential $U\left( x,t\right) $\ is an
analytical function of a variable $x$. For simplicity, in the
present work we only consider a harmonic oscillator with a
potential:
\begin{equation}
U\left( x\right) =\frac{kx^{2}}{2}.  \label{6}
\end{equation}
Let us introduce an exponential representation for the wave
functional:
\begin{equation}
\Psi \left[ x\left( t\right) \right] \equiv \exp \left(
\frac{i}{\widetilde{ \hbar }}S\left[ x\left( t\right) \right]
+\sigma \left[ x\left( t\right) \right] \right) ,
\label{7}
\end{equation}
where $S\left[ x\left( t\right) \right] ,\sigma \left[ x\left( t\right) %
\right] $ are unknown analytical functionals. The operator
$\widehat{I}$ is Hermitian with respect to the scalar product
(\ref{1}), therefore, its eigenvalues are real. From the equation
(\ref{2}), with account of (\ref{7}), we obtain an eigenvalue:
\begin{eqnarray}
\lambda &=&\int_{0}^{T}\left[ \overset{\cdot }{x}\frac{\delta
S}{\delta x}-
\frac{1}{2m}\left( \frac{\delta S}{\delta x}\right) ^{2}-U\right.  \notag \\
&&\left. +\frac{\widetilde{\hbar }^{2}}{2m}\left( \left(
\frac{\delta \sigma }{\delta x}\right) ^{2}+\frac{\delta ^{2}\sigma
}{\delta x^{2}}\right) \right] dt,
\label{8}
\end{eqnarray}%
and, in addition, a condition of its reality:
\begin{equation}
\int_{0}^{T}\left[ \overset{\cdot }{x}\frac{\delta \sigma }{\delta x}-\frac{1%
}{m}\frac{\delta S}{\delta x}\frac{\delta \sigma }{\delta x}\right. \left. -%
\frac{1}{2m}\frac{\delta ^{2}S}{\delta x^{2}}\right] dt=0.  \label{9}
\end{equation}
The representation (\ref{8}) is not final, because eigenvalues must
have no dependence on a trajectory $x\left( t\right) $ except for
boundary points $x_{0}=x\left( 0\right)$ and $x_{T}=x\left( T\right)
$. Let us introduce a functional series:
\begin{eqnarray}
S\left[ x\left( t\right) \right] &=&\int_{0}^{T}S_{1}\left( t\right) x\left(
t\right) dt  \label{10} \\
&&+\frac{1}{2}\int_{0}^{T}\int_{0}^{T}S_{2}\left( t,t^{\prime }\right)
x\left( t\right) x\left( t^{\prime }\right) dtdt^{\prime }+...,  \notag
\end{eqnarray}
\begin{eqnarray}
\sigma \left[ x\left( t\right) \right]  &=&\int_{0}^{T}\sigma _{1}\left(
t\right) x\left( t\right) dt  \label{11} \\
&&+\frac{1}{2}\int_{0}^{T}\int_{0}^{T}\sigma _{2}\left( t,t^{\prime }\right)
x\left( t\right) x\left( t^{\prime }\right) dtdt^{\prime }+....  \notag
\end{eqnarray}%
Then one obtains the final representation for the eigenvalue:
\begin{eqnarray}
\lambda &=&S\left\vert _{0}^{T}\right.
-\frac{1}{2m}\int_{0}^{T}S_{1}^{2}
\left( t\right) dt  \notag \\
&&+\frac{\widetilde{\hbar }^{2}}{2m}\int_{0}^{T}\left( \sigma _{1}^{2}\left(
t\right) +\sigma _{2}\left( t,t\right) \right) dt,  \label{12}
\end{eqnarray}
and, in addition to the reality condition (\ref{9}), the condition
of its independence on a trajectory $x\left( t\right) $:
\begin{eqnarray}
&&\int_{0}^{T}dt\left[ \overset{\cdot }{S}+\frac{1}{2m}\left( \frac{\delta S%
}{\delta x}\right) ^{2}+U\right.  \notag \\
&&\left. -\frac{\widetilde{\hbar }^{2}}{2m}\left( \left( \frac{\delta \sigma
}{\delta x}\right) ^{2}+\frac{\delta ^{2}\sigma }{\delta x^{2}}\right) %
\right]  \notag \\
&&+\frac{\widetilde{\hbar }^{2}}{2m}\int_{0}^{T}\left( \sigma _{1}^{2}\left(
t\right) +\sigma _{2}\left( t,t\right) \right) dt  \notag \\
&&-\frac{1}{2m}\int_{0}^{T}S_{1}^{2}\left( t\right) dt\left. =\right. 0.
\label{13}
\end{eqnarray}%
Here $\overset{\cdot }{S}$ denotes series (\ref{10}) in which the
coefficients are replaced by their time derivatives.

Now, we are ready to formulate the quantum action principle. First
of all, functional equations (\ref{9}) and (\ref{13}) must be
solved. They are equivalent to a set of the first-order differential
equations for coefficients of series (\ref{10}) and (\ref{11}). A
solution of this system depends on initial values of the
coefficients at the moment $t=0$. Therefore, the eigenvalue
$\lambda$, Eq.~(\ref{12}), of the action operator is a function of
the initial data. It is this function that has to be extremal in the
quantum action principle. In the next section this principle will be
considered in the simplest case of a harmonic oscillator.

\section{QUANTUM DYNAMICS OF HARMONIC OSCILLATOR}

In the case of harmonic oscillator it is enough to consider
quadratic terms in series (\ref{10}) and (\ref{11}), taking into
account that they are local with respect to the time variable, i.e.,
\begin{eqnarray}
S_{2}\left( t,t^{\prime }\right) &\equiv &S_{2}\left( t\right) \delta \left(
t-t^{\prime }\right) ,  \notag \\
\sigma _{2}\left( t,t^{\prime }\right) &\equiv &\sigma _{2}\left( t\right)
\delta \left( t-t^{\prime }\right) .  \label{14}
\end{eqnarray}
In this case, functional equations (\ref{9}) and (\ref{13}) are
equivalent to the following set of differential equations:
\begin{eqnarray}
&&\overset{\cdot }{\sigma }_{1}+\frac{1}{m}\left( \sigma _{1}S_{2}+\sigma
_{2}S_{1}\right) \left. =\right. 0,  \notag \\
&&\overset{\cdot }{\sigma }_{2}+\frac{1}{m}\sigma _{2}S_{2}\left. =\right. 0,
\notag \\
&&\overset{\cdot
}{S}_{1}+\frac{1}{m}S_{1}S_{2}-\frac{\widetilde{\hbar }^{2}
}{2m}\sigma _{1}\sigma _{2}\left. =\right. 0,  \notag \\
&&\overset{\cdot
}{S}_{2}+\frac{1}{m}S_{2}^{2}+k-\frac{\widetilde{\hbar }^{2}
}{m}\sigma _{2}^{2}\left. =\right. 0,
\label{15}
\end{eqnarray}
Moreover there is the following algebraic equation for the initial
data:
\begin{eqnarray}
&&\left. \left( \sigma _{1}x+\sigma _{2}\frac{x^{2}}{2}\right) \right\vert
_{0}^{T}  \notag \\
&&-\frac{1}{m}\int_{0}^{T}\left( \sigma _{1}S_{1}+2S_{2}\right) dt\left.
=\right. 0.  \label{16}
\end{eqnarray}
A final representation of the action eigenvalue takes a form:
\begin{eqnarray}
\lambda &=&\left. \left( S_{1}x+S_{2}\frac{x^{2}}{2}\right) \right\vert
_{0}^{T}  \notag \\
&&-\frac{1}{2m}\int_{0}^{T}S_{1}^{2}dt+\frac{\widetilde{\hbar
}^{2}}{2m} \int_{0}^{T}\left( \sigma _{1}^{2}+\sigma _{2}\right) dt.
\label{17}
\end{eqnarray}
Exponential representation (\ref{7}) of the wave functional is the
most suitable for the formulation of quasi-classical asymptotics.

Let us consider the classical limit $\left( \widetilde{\hbar
}=0\right) $\. In this limit, from Eq.~(\ref{15}) one can obtain:
\begin{eqnarray}
S_{1} &=&S_{10}\frac{\cos \omega _{0}t_{0}}{\cos \omega _{0}\left(
t-t_{0}\right) },  \notag \\
S_{2} &=&-\sqrt{mk}tg\omega _{0}\left( t-t_{0}\right) ,  \label{18}
\end{eqnarray}
where $\omega _{0}\equiv \sqrt{k/m}$ is the own frequency of the
oscillator, $S_{10}$ is an initial value of the function
$S_{1}\left( t\right) $. The constant $t_{0}$ is related with the
initial value of the function $ S_{2}\left( t\right) $:
\begin{equation}
S_{20}=\sqrt{mk}tg\omega _{0}t_{0}.  \label{19}
\end{equation}
Substituting solution (\ref{18}) into Eq.~(\ref{17}), in the
classical limit we obtain:
\begin{eqnarray}
\lambda &=&S_{10}\left( x_{T}\frac{\cos \omega _{0}t_{0}}{\cos \omega
_{0}\left( T-t_{0}\right) }-x_{0}\right)  \notag \\
&&-\frac{1}{2}\sqrt{mk}\left( x_{T}^{2}tg\omega _{0}\left( T-t_{0}\right)
+x_{0}^{2}tg\omega _{0}t_{0}\right)  \notag \\
&&-S_{10}^{2}\frac{\cos ^{2}\omega _{0}t_{0}}{2\sqrt{mk}}\left( tg\omega
_{0}\left( T-t_{0}\right) +tg\omega _{0}t_{0}\right) .  \label{20}
\end{eqnarray}
Now we must find an extremum of this function with respect to two
variables, $S_{10}$ and $t_{0}$. The extremum condition gives for
$S_{10}$:
\begin{equation}
S_{10}=\sqrt{mk}\frac{x_{T}\cos \omega _{0}t_{0}-x_{0}\cos \omega
_{0}\left( T-t_{0}\right) }{\cos \omega _{0}t_{0}\sin \omega _{0}T}.
\label{21}
\end{equation}
However, the constant $t_{0}$ remains indefinite. Therefore, the
eigenvalue $\lambda $ is degenerate in the classical limit. As it
must be in the classical limit, this eigenvalue is equal to the
classical action of the harmonic oscillator (see, for example,
\cite{F}):
\begin{equation}
\lambda =\sqrt{mk}\frac{\left( x_{T}^{2}+x_{0}^{2}\right) \cos \omega
_{0}T-2x_{T}x_{0}}{2\sin \omega _{0}T}.  \label{22}
\end{equation}
The corresponding eigenfunctional has a phase which is parameterized
as follows:
\begin{equation}
S\left[ x\left( t\right) \right] =-\frac{\sqrt{mk}}{2}\int_{0}^{T}\left(
x\left( t\right) -\widetilde{x}\left( t\right) \right) ^{2}dt,  \label{23}
\end{equation}
where
\
\begin{equation}
\widetilde{x}\left( t\right) \equiv \frac{x_{T}\cos \omega
_{0}t_{0}-x_{0}\cos \omega _{0}\left( T-t_{0}\right) }{\sin \omega _{0}T\sin
\omega _{0}\left( t-t_{0}\right) }.  \label{24}
\end{equation}%
The parameter $\sigma \left[ x\left( t\right) \right] $ remains
indefinite in the classical limit. It is necessary to outline that
similar to ordinary quantum mechanics (see, for example,
\cite{Bohm}), the classical limit in our theory is valid up to the
first order in the constant $\widetilde{\hbar }$. The arbitrary
parameter $t_{0}$ can be interpreted as a degree of the oscillator
excitation. It plays a role of an energy parameter within our
theory.

\section{CONCLUSIONS}

We conclude that the new form of quantum mechanics has a proper
classical limit. Quantum corrections to the quantum action can give
essential predictions of the
new quantum theory. In a further paper, we will give a relationship between a new $\widetilde{%
\hbar }$ and the ordinary $\hbar $ Plank constants.

We are thanks A. V. Goltsev for useful discussions.




\end{document}